\documentclass[%
 reprint,longbibliography,preprintnumbers,
nofootinbib,
 amsmath,amssymb,
 aps,
pre,
]{revtex4-1}
\pdfoutput=1
\usepackage[utf8]{inputenc}
\usepackage{flushend}
\usepackage{dcolumn}% Align table columns on decimal point
\usepackage{bm}% bold math
\usepackage{balance}
\usepackage[normalem]{ulem}

\usepackage[colorlinks = true,
            linkcolor = blue,
            urlcolor  = blue,
            citecolor =green,
            anchorcolor = blue]{hyperref}
\usepackage{verbatim}
\usepackage{color,ulem}
\usepackage[english]{babel}

\usepackage[utf8]{inputenc}
\input Starburst.fd
\newcommand*\initfamily{\usefont{U}{Starburst}{xl}{n}}\initfamily

\newcommand{\beq}{\begin{eqnarray}}
\newcommand{\eeq}{\end{eqnarray}}
\usepackage{amsmath}
\usepackage{tikz}
\usetikzlibrary{decorations.pathmorphing}
\usetikzlibrary{shapes.misc}
\tikzset{cross/.style={cross out, draw=black, minimum size=8*(#1-\pgflinewidth), inner sep=0pt, outer sep=0pt},
cross/.default={1pt}}
\usetikzlibrary{patterns,math}
\begin{document}

\title{Fragility and thermal expansion control crystal melting and the glass transition}

\author{Alessio Zaccone$^{1,2}$}
%\author{Alois Loidl$^{3}$}
%\author{Peter Lunkenheimer$^{3}$}
\author{Konrad Samwer$^2$}

\address{$^1$  Department of Physics ``A. Pontremoli'', University of Milan, via Celoria 16,
20133 Milan, Italy}
\address{$^2$ 
I. Physikalisches Institut, University of Goettingen, Goettingen, Germany}
%\address{$^3$ 
%Experimental Physics V, Center for Electronic Correlations and Magnetism, University of Augsburg, Augsburg, Germany}

 \vspace{1cm}

\begin{abstract}
Analytical relations for the glass transition temperature, $T_g$, and the crystal melting temperature, $T_m$, are developed on the basis of nonaffine lattice dynamics. The proposed relations explain: (i) the seemingly universal factor of $\approx 2/3$ difference between glass transition temperature and melting temperature of the corresponding crystal, and (ii) the recent empirical discovery that both $T_g$ and $T_m$ are proportional to the liquid fragility $m$ divided by the thermal expansion coefficient $\alpha$ of the solid.
\end{abstract}

\maketitle

\section{Introduction}
Being able to predict the temperatures at which solids melt into liquids is one of the overarching goals of physics \cite{Zaccone_book,Chaikin,Born_Huang,Wallace}, with important ramifications for amorphous materials, where this temperature coincides with the glass transition temperature \cite{kob_book,Lunkenheimer2023}. For example, for glassy polystyrene, one of the most studied amorphous polymers, the melting of the amorphous solid into the liquid as signaled by a dramatic drop of the low-frequency shear modulus, occurs at $T=383$ K \cite{Zaccone_2013}, which is almost within the error bar of the glass transition temperature as measured by calorimetry, $T_g=380.15\pm2$ \cite{Rieger1996}. Similarly, in crystals, the thermodynamic melting temperature $T_m$ coincides with a sharp drop of the shear elastic constant $C_{44}$ (low-frequency shear modulus) \cite{Ma}. Indeed, for the melting of the FCC Lennard-Jones (LJ) crystal, this happens at $T_m=0.79$ \cite{Ma}, in units of $\epsilon/k_{B}$, where $\epsilon$ is the energy scale of the LJ potential and $k_B$ the Boltzmann's constant.
In the case of 2D melting, the Berezinskii-Kosterlitz-Thouless (BKT) theory offers a full mechanistic understanding \cite{Kosterlitz_1973} and the transition from a rigid solid to a liquid with zero rigidity occurs via an intermediate hexatic phase. In comparison, crystal melting in 3D has remained much less understood \cite{Ma,Tallon1989,Kostya,Zaccone_Weitz,Hirshberg,Douglas_melt,deWith}.

The Lindemann melting rule \cite{Lindemann} stipulates that a solid melts when the mean atomic root-mean-square displacement becomes equal to approximately $0.1$ times the nearest-neighbor distance, which is roughly the distance of the inflection point in Lennard-Jones type interatomic potentials. However, there is a significant variability of this number across different materials and chemistries. 
The Born melting rule, on the other hand, stipulates that a solid ceases to be a solid when it starts to mechanically behave like a liquid, i.e. when its zero-frequency shear modulus becomes equal to zero \cite{Born_melting,Born_Huang}.

There are currently no theories that can predict the melting temperatures of 3D crystals without adjustable parameters \cite{deWith}. The Born criterion applied to amorphous materials, on the other hand, has been comparatively successful in providing compact expressions for the glass transition temperature, $T_g$, in reasonable parameter-free agreement with the experimental data \cite{ZacconePRL} 
(where the only adjustable parameter in the comparison with the experimental data is the glass transition itself, while all the other parameters are determined by the molecular chemistry of the specific material, including the thermal expansion coefficient which accounts for anharmonicity). There are, however, reports of elemental crystals violating the Born criterion \cite{RMP_inst}, based on the fact that the low-temperature $T$-dependence of the shear elastic constant does not extrapolate smoothly to zero at melting (but rather goes through a sharp drop at melting, as shown in \cite{Ma}). There is no real "violation", however, if one considers that it is this sharp drop (governed by the rise of the negative nonaffine correction to the shear modulus, due to thermal fluctuations and defects proliferation) which marks the melting phenomenon, and not the extrapolation of the low-temperature $T$-dependence of the modulus to zero. The  "violation" is often associated with the presence of imaginary frequencies, also known as instantaneous normal modes (INMs) \cite{Douglas_melt}. Typically, however, a full lattice dynamic calculation of elastic moduli of a real solid must also include the INMs in order to quantitatively reproduce experimental or simulated values, as shown in the context of amorphous materials \cite{Palyulin,Elder,Douglas_INMs,Vaibhav}.

The Born criterion can be summarized as $G(T_c)=0$, where $T_c$ is either the melting temperature or the glass transition temperature, and $G$ the zero-frequency shear modulus. While elasticity theory as we know it from continuum mechanics has no length scale in it, modern lattice dynamics provides the atomic-scale mathematical framework to predict elastic constant values starting from the atomic structure and from the interatomic interactions \cite{Cui2019}. Due to either intrinsically non-centrosymmetric environment of atomic structure, as in glasses and in non-centrosymmetric crystals, or due to instantaneous breaking of centrosymmetry due to thermal motions, besides the so-called affine contribution to the shear elastic constant, which is just the sum of stretching energy of bonds in the direction of strain, there exists a ubiquitous negative contribution (hence, causing softening) which arises from the nonaffine displacements, i.e. the atomic displacements that are performed on top of the affine displacements dictated by the macroscopic strain tensor. Hence, $G=G_A - G_{NA}$, where $G_A$ is the affine or Born or infinite-frequency modulus, and $-G_{NA}$ is the nonaffine contribution.
Furthermore, the nonaffine contribution is important also in the vicinity of dislocations where the inversion symmetry of the lattice is broken by the topological defect.

From the point of view of Born melting, one can thus attempt a unified description of crystal melting and of the glass transition in terms of the $G=G_A - G_{NA}=0$ rule. However, such a framework should also be able to explain the famous 2/3 ratio rule, i.e. $T_g \approx (2/3)T_m$ \cite{Angell_1986,Uitert,Lunk_2024}. At present, there is no theoretical derivation for this empirical rule, which is nevertheless supported by a large amount of experimental data \cite{Lunk_2024}. The only attempt, to our knowledge, is that of Kanno \cite{Kanno} who, however, had to postulate a root mean square ratio of displacement in glass over displacement in crystal equal to 0.9 (which is incompatible with most simulations \cite{Rongchao,Koun}), in an ad-hoc fashion, to be able to retrieve the 2/3 ratio.

Finally, recent analysis carried out for nearly a hundred different materials \cite{Lunkenheimer2023,Lunk_2024}, has highlighted the existence of universal relations between the glass transition and melting temperatures and the liquid fragility $m$ and solid thermal expansion coefficient $\alpha$:
\begin{align}
    T_g & \propto m/\alpha, \\
     T_m & \propto m/\alpha, \label{rel}
\end{align}
for which no theoretical derivation is available.
We recall that the fragility $m$ indicates the degree of non-Arrhenius temperature dependence displayed by the viscosity of a supercooled liquids. Fragile liquids have a large value of $m$, which means that the temperature-dependent increase of viscosity upon reducing the temperature close to $T_g$ is qualitatively different compared to that predicted by Arrhenius' law (typically, a double exponential function of temperature \cite{KSZ}. Strong liquids, on the other hand, have low values of $m$ and the strongest among them follow the Arrhenius law all the way to vitrification \cite{Angell,KSZ}.
Fragility is also a measure of the cooperativity in the dynamics of the liquid, with Adam-Gibbs type theories suggesting that fragility specifies how fast the size of CRRs changes with temperature \cite{Wolynes}. Within the KSZ model, which attempts to relate fragility $m$ to the details of the potential of mean force between atoms and molecules, the fragility is proportional to the steepness parameter $\lambda$, which quantifies the steepness of repulsion in the potential of mean force \cite{KSZ,Lunk_2020}.

In this paper, we will first provide a derivation of the approximately 2/3 ratio rule $T_g \approx (2/3)T_m$ using the Lindemann approach first and the Born criterion afterwards, which lead to the same result.
Then we will offer a mathematical derivation of Eqs. \eqref{rel} based on the nonaffine lattice dynamics theory \cite{Zaccone_book,Scossa,Cui2019}.

\section{2/3 ratio rule from the Lindemann criterion}
It is well known that the energy required to ``melt'' a glass, i.e. to bring an amorphous solid into the liquid state by heating, is a factor $2/3$ of the energy required to melt the corresponding crystalline state \cite{Angell_1986,Ranko_2006,Granato,Kanno}. 
The seemingly universal relation can be written as:
\begin{equation}
\frac{k_B T_g}{k_B T_m} \approx \frac{2}{3}
\label{crystal_glass}
\end{equation}
where $T_g$ denotes the glass transition temperature and $T_m$ the melting temperature of the crystal. While the melting temperature is thermodynamically defined, the glass transition temperature can be defined in various ways depending on the chosen observable, and different definitions exist in the literature. Experimentally, it is operationally (and conventionally) defined as the temperature at which the relaxation time $\tau$ is equal to $100$ s. Although this definition is widely used in the experimental practice, it is not convenient for theoretical studies, because of its arbitrariness and of its lacking a mechanistic rationale. In the following, we shall more precisely define the glass transition temperature $T_g$ as the temperature at which the solid-liquid transition occurs, i.e. where the amorphous solid starts to flow like a liquid. In turn, the solid-liquid (rigidity) transition coincides with the temperature at which the low-frequency (plateau) shear modulus becomes very small (otherwise the solid would not be able to ''flow") \cite{Zaccone_2013}. From an experimental point of view, since no experiment can be carried out at exactly zero-frequency, the frequency should be $1/(2 \pi 100)$ Hz, which corresponds to the conventional criterion of declaring $T_g$ as the temperature at which the relaxation time is $\tau=100$ s.

With this definition, we shall be able to determine the $T_g$ as a function of well-defined and quantifiable physical quantities related to bonding, molecular packing density etc.

Furthermore, the $T_g$ and the $T_m$ separately obey an inverse proportionality relation with the thermal expansion coefficient, in the glass and crystal phase respectively \cite{ZacconePRL,Granato,Lunkenheimer2023},
\begin{align}
T_g &= \frac{c_{g}}{\alpha_g} \nonumber\\
T_m &= \frac{c_{x}}{\alpha_x}
\label{constant}
\end{align}
where $c_{g}$ is a constant that also includes the fragility $m$ of the corresponding liquid as shown in Ref. \cite{Lunkenheimer2023}. In the context of glasses, the inverse proportionality between $T_g$ and thermal expansion $\alpha_g$ (or in terms of the liquid thermal expansion, $\alpha_l$ ,which is, anyway, directly proportional to $\alpha_g$ \cite{Lunkenheimer2023}) is known as the Simha-Boyer rule.
The energy required to melt the solid, according to the Lindemann criterion, is proportional to the internal bonding energy, via the Debye temperature, which is related to the bond spring constant via the Debye frequency. 
For a 3D solid the Lindemann criterion gives \cite{Granato}
\begin{equation}
k_B T_m \sim \Theta_{D,x}^{2} \, M v^{2/3}
\label{melting}
\end{equation}
where $\Theta_{D,x}$ is the Debye temperature of the crystal, $M$ is the atomic mass and $v$ is the volume per atom. 
A similar Lindemann criterion applies to atomic glasses as well \cite{SamwerPRL2015} so that we can write \begin{equation}
k_B T_g \sim \Theta_{D,g}^{2} \, M v^{2/3}
\label{glass}
\end{equation}
also for the glass transition, where $\Theta_{D,g}$ is the Debye temperature of the glass.

Furthermore, for an isotropic 3D solid, we have that the Debye temperature is given by:
\begin{equation}
\Theta_D = \frac{\hbar }{k_B} \omega_D
\label{Debye_temp}
\end{equation}
where $\omega_D$ is the Debye frequency.
Also, for an isotropic 3D solid we have:
\begin{equation}
\omega_D = \sqrt[3]{\frac{6 \pi^2 N}{V}}c_{s}
\label{Debye_freq}
\end{equation}
where $c_{s}$ is the longitudinal speed of sound, which is always larger than the transverse speed of sound. Near melting or glass transition the bulk modulus $K$ is much larger than the shear modulus $G$ because the latter is vanishing at the transition (Born melting) whereas the bulk modulus remain large, hence $K \gg G$. Also in general, the bulk modulus is always larger than the shear modulus because of lower nonaffine contributions as mathematically demonstrated in \cite{JAP,Schlegel2016}.
Hence it is legitimate to assume, for the longitudinal speed of sound near melting:
\begin{equation}
c_{s} = \sqrt{\frac{K + \frac{4}{3}G}{\rho}}\approx \sqrt{\frac{K}{\rho}}
\label{sound}
\end{equation}
where $\rho$ is the atomic density.

According to microscopic (Born-Huang) affine elasticity theory, the bulk modulus is given by the affine deformation theory, for both crystalline and disordered systems \cite{JAP,Schlegel2016} as
\begin{equation}
K \sim \kappa z
\label{bulk}
\end{equation}
where $\kappa$ is the bond spring constant and $z$ is the coordination number or mean number of nearest neighbours \cite{Phillips_Thorpe}. 

By combining Eqs. \eqref{Debye_freq}, \eqref{sound}, \eqref{bulk} into Eq. \eqref{Debye_temp}, we thus obtain:
\begin{equation}
\Theta_D \sim \sqrt{\kappa z}.
\end{equation}

Inserting this result into Eqs. \eqref{melting}-\eqref{glass}, we thus obtain:
\begin{align}
k_B T_g &\sim \kappa \, z_g \nonumber \\
k_B T_m &\sim \kappa \, z_x,
\label{spring}
\end{align}
where the bonding spring constant $\kappa$ can be assumed to be the same for glass and crystal, whereas $z_g$ and $z_x$ denote the interparticle connectivity in the glass and in the crystal, respectively.

We should note that experimental and simulations data for the radial distribution function (rdf) give a first coordination peak which is roughly comparable in the glass and in the crystal (only the long-range order is different). The rdf in the glass is, of course, broader, but not all the neighbors which contribute to the rdf are relevant for the mechanical response. Indeed, some of the neighbors included in the first peak of the rdf are highly fluctuating in and out of contact, and therefore do not significantly contribute to the elasticity \cite{Paddy}.

This is because the rdf is a only an average snapshot of the structure, and, moreover, it does not change significantly upon crossing the glass transition from liquid to glass \cite{Donth,kob_book}.
This is, indeed, because it does not take into account that, in the incipient glass at the glass transition, only a subset of the atoms in the first coordination shell are long-lived and contribute to the elastic properties. This has been demonstrated experimentally for the case of colloidal glass by Laurati, Egelhaaf and co-workers \cite{Egelhaaf_2017}. Hence, above and in the following, the connectivity at the glass transition $z_g$ does not refer to the coordination number computed from the integral of the first peak of the rdf \cite{Hansen}, but to a smaller number that represents the mean number of long-lived, \emph{mechanically stable} interparticle contacts. 

For the glass, assuming that the building blocks are spherical-like atoms or molecules interacting via central-force potentials, one can approximate the structure as a random close packing with $z_g \approx 6$ \cite{Scossa,Zaccone_book}. This estimate remains approximately valid for molecular liquids interacting via non-covalent bonds. For molecules with aspect ratio larger than 1, $z_g$ will be larger than 6, and approaching a maximum of 10 for rod-like molecules \cite{Zaccone_book}.

Hence, from Eq. \eqref{spring} we obtain the fundamental relationship between the glass temperature to melting temperature ratio and the respective atomic connectivities in the incipient glass and crystal states, respectively:
\begin{equation}
    \frac{T_g}{T_m}=\frac{z_g}{z_x}.\label{ratio}
\end{equation}

According to the Alexander-McTague argument \cite{McTague}, the bcc crystal phase is favoured near the melting line, and as observed for many experimental systems \cite{Zaccone_Weitz,Jiang}, we thus have 
$z_{m}=z_{x,bcc}= 8$.

By using this assumption in Eq.\eqref{ratio}, we obtain the following estimate:
\begin{equation}
     \frac{T_g}{T_m}=\frac{z_g}{z_{x,bcc}}=\frac{6}{8}=0.75
    \label{connectivity1}
\end{equation}
which is somewhat larger than the most quoted value $2/3 \approx 0.667$ \cite{Angell_1986, Ranko_2006,Granato}.

If, instead, it is the fcc crystal phase the one encountered upon crystallization from the melt, we have $z_{m}=z_{x,bcc}= 12$ and
\begin{equation}
      \frac{T_g}{T_m}=\frac{z_g}{z_{x,fcc}}=\frac{1}{2}.
    \label{connectivity2}
\end{equation}
Hence we have obtained an upper bound $\approx 0.75$ and a lower bound $\approx 0.5$, which are in agreement with the observation that the experimental data for various materials roughly vary between $0.5$ and $0.8$ \cite{Novikov_quantum,Poeschl,Lunk_2024}.

This allows us to constrain the $T_g/T_m$ ratio in the following interval:
\begin{equation}
    0.5 \leq \frac{T_g}{T_m} \leq 0.75.
\end{equation}

Upon assuming that fcc and bcc, on average for a large variety of different systems, are equally probable at melting, we obtain the following arithmetic average:
\begin{equation}
    \langle \frac{T_g}{T_m} \rangle = 0.625
\end{equation} 
which is not far from the most quoted value $0.65$ found from fitting a large body of data from different experimental systems \cite{Lunk_2024}.

For polymers, the critical average connectivity at the glass transition in the amorphous state was estimated in Ref. \cite{ZacconePRL}, as $z_g = 4$, which comes from considering both central-force interactions due to van der Waals interaction between monomers (for which rigidity is achieved when $z=6$) and covalent bonds along the chain which have angular bond-bending constraints (for which rigidity is achieved when $z=2.4$ \cite{Zaccone_2013}).
In the polymer crystalline state, as famously shown by Keller in 1957 \cite{Keller}, the crystal structure is given by the folded-chain stackings. In this structure, each monomer has contacts with $6$ neighbours: two covalently bonded neighbours along the chain, two non-covalently interacting monomers belonging to the same (folded) chain each separated by one fold (in the same plane as the folded chain), and two more non-covalently interacting monomers belonging to the previous and to the next chain in the same stack (in the plane orthogonal to the folded chain).
This gives $z_{x,pol}=6$. Hence, for polymers we have:
\begin{equation}
      \frac{T_g}{T_m}=\frac{z_{g,pol}}{z_{x,pol}}=\frac{4}{6}=\frac{2}{3}
\end{equation}
which exactly coincides with the most quoted value $2/3$ \cite{Angell_1986, Ranko_2006,Granato}.

Finally, we should also consider the case of network oxides and ceramics, such as e.g. silica glass. These are random networks that approximately follow Zachariasen's model \cite{Zachariasen} (each oxygen is bonded to no more than two forming cations and each cations to $\leq 4$ other atoms). In silica, each oxygen is bonded to two Si atoms, and each Si atom is bonded to 4 oxygens. By accounting for the stoichiometry, the average coordination in $\beta$-quartz is equal to $z=2.6667$. At the glass transition, as the silica melt becomes rigid, the shear modulus vanishes at $z=2.4$ as found by different methods \cite{He_Thorpe,Zaccone_2013}. Hence, $z_g=2.4$ and $z_x=2.6667$, which gives $T_g/T_m = 0.8998$, significantly larger than the most quoted value $0.6667$.
However, if one considers the Si atoms to be the effective elastic building blocks, and each Si atom to be interacting with 4 nearest-neighbour Si atoms via an ''effective" bond given by the Si-O-Si connection, the situation is different. Then one has again $z_g=2.4$ because that is the universal value at which rigidity is achieved right at the glass transition \cite{Phillips_Thorpe,He_Thorpe}, whereas for $\beta$ quartz one has $z_x=4$ since that is the mechanical coordination of each Si atom. In that case, we get $T_g/T_m = 2.4/4=0.6$, which is again not too far from the most quoted value $2/3$.

Hence, by using the Lindemann or Born melting criterion we obtain via the above the argument that monoatomic elemental systems and molecular organic systems have $T_g/T_m=0.625$, network oxides have $T_g/T_m=0.6$, and polymers $T_g/T_m=2/3$. This model thus provides a microscopic insight and justification to the "universal" $2/3$ rule often quoted for the $T_g/T_m$ ratio across chemically different systems.

\section{2/3 ratio rule from the Born criterion}
The above results about the $2/3$ ratio between glass transition and melting temperatures can also be obtained directly from the Born melting criterion by taking nonaffinity of atomic displacements into account. Within the nonaffine response theory \cite{Zaccone_book} valid for both glasses and crystals at finite temperature, the shear modulus is given by $G(T)=G_A(T) - G_{NA}(T)$ where $G_A$ is the affine (Born-Huang) contribution and $ - G_{NA}(T)$ is the nonaffine correction. Both depend on temperature $T$. As shown by different authors using numerical calculations of the nonaffine response theory \cite{chinese_nonaffine,ilg}, the nonaffine part is an approximately linearly increasing function of $T$, i.e. $-G_{NA}(T)=-b T$. At both melting and glass transition, according to the Born melting criterion, $G(T)=G_A(T) - G_{NA}(T)=0$, which will occur at a higher temperature for the crystal given the higher connectivity of the crystal with respect to the glass. Therefore we can write:
\begin{equation}
    G_{x}=c z_x -bT_m = G_{g}=c z_g -b T_g =0
\end{equation}
from which it follows that $c=b T_g/z_g$. Upon substituting this relation in $c z_x -bT_m =0$ we then get:
\begin{equation}
\frac{T_g}{T_m}=\frac{z_g}{z_x}
\end{equation}
which is the same result that we obtained above in Eq. \eqref{spring} using a different route, i.e. the Lindemann criterion. Hence, both the Lindemann and the Born criterion lead to the same result for the ratio between glass transition and melting temperature. 
Importantly, the two derivations demonstrate that the $\frac{T_g}{T_m}$ ratio depends uniquely on the atomic connectivity in glass and crystal. It does not depend on other factors such as the different thermal expansion in the liquid and in the solid or the fragility. This is also because the thermal expansion coefficient of the glass and of the crystal have, typically, very similar values, $\alpha_g \approx \alpha_x$, and both are, on average, about one third of the thermal expansion coefficient of the corresponding liquid, $\alpha_g \approx \alpha_x \approx \alpha_l /3$ \cite{Lunkenheimer2023}.

Finally, we should remark on the fact that the above Born melting scenario, and the corresponding estimates, are also fully consistent with an increasing number of point defects (vacancies, dislocations) as the melting is approached from below. This is because all point defects break the local inversion symmetry of the lattice, which gives rise to important (negative) nonaffine contributions to the shear modulus \cite{Johannes,Milkus,Zaccone_book}. Hence, the local breaking of centrosymmetry due to defects directly contributes to $-G_{NA}$ in the above equations, along with the thermal fluctuations. 

We can, therefore, identify the following mechanistic effects responsible for the decrease of the zero-frequency shear modulus upon increasing the temperature towards the melting temperature: (i) proliferation of vacancies and dislocations, which have the main effect of making $-G_{NA}$ more important, hence lowering the overall $G$ as $T$ rises; (ii) thermal fluctuations, which bring nearest-neighbours across the inflection point in the interatomic potential, which also contribute to making $-G_{NA}$ more important, hence lowering the overall $G$ as $T$ rises. In future work, more quantitative estimates of the relative importance of effects (i) and (ii) may be extracted from atomistic simulations for specific systems.

\section{$T_g$, $T_m$ as functions of fragility and thermal expansion}
It has been shown in Ref. \cite{Lunkenheimer2023}, by collecting experimental data on tens of different glass-forming liquids, that the following universal relationship:
\begin{equation}
    T_g \propto m/\alpha \label{nat_phys}
\end{equation}
is the only relation that can describe the entire data set. In this relation, $m$ is the fragility index \cite{Angell}, defined as:
\begin{equation}
m \equiv \frac{T_g}{\ln 10}\left( -\frac{\partial{\ln \eta }}{\partial T } \right) _{T=T_{g}} \label{fragility_def}
\end{equation}
More recently in \cite{Lunk_2024}, based on a set of experimental data from the literature, and arguing also on the fact that Eq. \eqref{nat_phys} is incompatible with $T_m \sim \/\alpha$ and the $2/3$ rule, it was shown that the correct relationship for the melting temperature must be:
\begin{equation}
    T_m \propto m/\alpha \label{Lunk}
\end{equation}
where $\alpha$ is here the crystal thermal expansion coefficient. 

\begin{figure}
\includegraphics[width = 1.0 \linewidth]{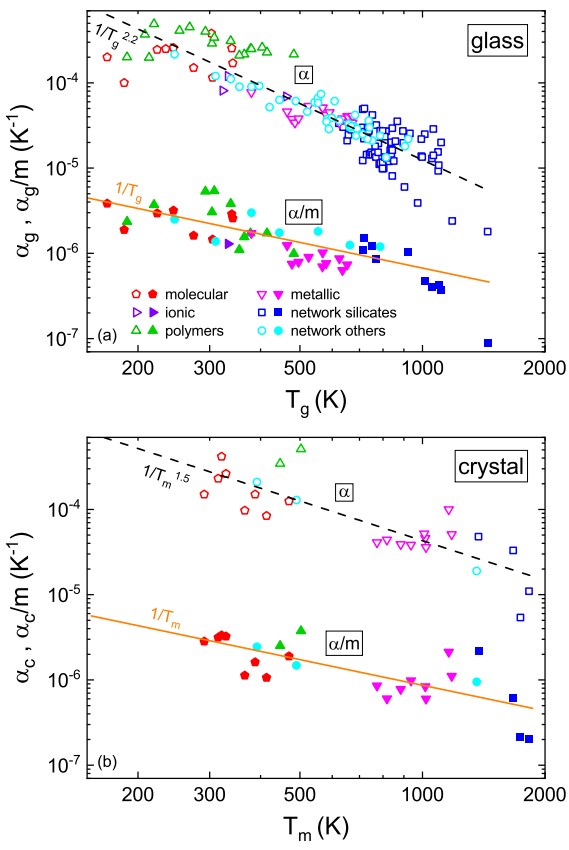}
	\caption{Volume thermal expansion coefficients $\alpha_g$ in the glass
state (a) and $\alpha_c$ in the crystalline phase (b) plotted double-logarithmically vs $T_g$ or $T_m$, respectively. The figure is reproduced with permission from Ref. \cite{Lunk_2024}. The data cover a large
number of materials belonging to different material classes as indicated in the
legend. The open symbols represent the bare expansion coefficients
while the closed symbols show $\alpha$ divided by the fragility $m$. The dashed lines
are power-law fits, of the bare $\alpha$ data (open symbols), leading to
exponents of about $-2.2$ for $\alpha_g$ and $-1.5$ for $\alpha_c$. The solid lines are
fits of the $\alpha/m$ data with slope -1. Adapted with permission of the American Physical Society from Ref. \cite{Lunk_2024}.}
		\label{fig1}
\end{figure}

This new relation, Eq. \eqref{Lunk}, is the best fit to the known experimental data across various systems, as shown in Fig. \ref{fig1} (reproduced from \cite{Lunk_2024}).

For both these relations, Eq. \eqref{nat_phys} and Eq. \eqref{Lunk}, there is currently no theoretical derivation. We shall derive them here using the extended Born melting criterion that accounts also for nonaffine deformations \cite{ZacconePRL}.

Let us start once again from the Born melting criterion, written in terms of the modern microscopic elasticity theory, which takes nonaffinity into account \cite{Slonczewski,Lemaitre,chinese_nonaffine,Scossa,wittmer}. According to the theory (cfr. Chapter 2 in Ref. \cite{Zaccone_book}), valid for both crystals and glasses, the solid-liquid transition occurs when the static (zero-frequency) shear modulus of the solid state vanishes:
\begin{equation}
    G(T_c) = G_A(T_c) - G_{NA}(T_c)=0
\end{equation}
where $T_c$ is the temperature of the transition, i.e. either glass or melting. This equation could be also interesting in the context of the shoving model of the glass transition, where a debate is ongoing about whether the activation energy for shoving events should be identified with the high-frequency shear modulus \cite{Dyre,Dyre2} or rather with an intermediate-frequency shear modulus as suggested by Leporini, Douglas and co-workers \cite{Puosi,Xu2023}.
We thus get, at the transition temperature:
\begin{equation}
    G_A(T_c) \approx G_{NA}(T_c) = const. \label{tran}
\end{equation}
Since the affine modulus coincides with the infinite-frequency modulus \cite{Zwanzig}, $G_A \equiv G_{\infty}$, we thus have the Maxwell relaxation time $\tau_M$ given by:
\begin{equation}
    \tau_M = \frac{\eta}{G_A}.
\end{equation}
Right at the transition, by virtue of Eq. \eqref{tran}, we, therefore, also have:
\begin{equation}
    \tau_M G_{NA} = \eta. \label{Max}
\end{equation}
We can now take the logarithm of both sides in Eq. \eqref{Max} and then multiply both sides by the factor $-\frac{T_g}{ \ln 10 }\frac{\partial}{\partial T}$, resulting in:
\begin{equation}
    \frac{T_g}{\ln 10}\left[ - \frac{\partial}{\partial T}\ln (G_{NA}(T_c) \tau_M )\right]_{T_c} =m
\end{equation}
which, using the well known properties of the logarithm, becomes:
\begin{equation}
    \frac{T_g}{\ln 10}\left[ - \frac{\partial}{\partial T}(\ln const  + \ln \tau_M )\right]_{T_c} =m,
\end{equation}
and hence
\begin{equation}
    \frac{T_g}{\ln 10}\left[ - \frac{\partial}{\partial T}(\ln \tau_M )\right]_{T_c} =m,
\end{equation}
and, finally,
\begin{equation}
    \frac{T_g}{\ln 10}\left[ -\frac{1}{\tau_M} \frac{\partial \tau_M} {\partial T}\right]_{T_c} =m.
\end{equation}
We shall recall that the quantities on the left hand side of the above relation are to be evaluated at $T_c$ or in its vicinity.

For all known liquids, the Maxwell time is a monotonically decreasing function of $T$, both near freezing as well as near the glass transition \cite{Egami_1,Egami_2}. Hence,
\begin{equation}
    \left[ -\frac{1}{\tau_M} \frac{\partial \tau_M} {\partial T}\right]_{T_c}  = c  > 0.
\end{equation}
Of course, one could argue that the coefficient $c$ is also fragility-dependent. However, it should be noted that the Maxwell time $\tau_M$ typically has a significantly weaker temperature dependence compared to the $\alpha$-relaxation time or the viscosity \cite{Egami_1,Egami_2} (as a matter of fact, it is defined as viscosity divided by high-frequency shear modulus, the latter also having some temperature dependence). All in all, we expect the fragility-dependence of the coefficient $c$ to be weak. This assumption is \emph{a posteriori} confirmed by its ability to recover the universal relation between fragility, glass transition, and thermal expansion, observed for a great number of glass-forming materials \cite{Lunkenheimer2023}.
To summarize, we just obtained the following relations, valid for both crystal melting and glass transition:
\begin{align}
    T_g &= \left(\frac{\ln 10}{c\, G_{NA}(T_g)}\right) m,\label{ket} \\
     T_m &= \left(\frac{\ln 10}{c'\, G_{NA}(T_g)}\right) m 
\end{align}
where, in general, we expect $c' \neq c = const$. Since, in the above expressions, everything that multiplies $m$ is a constant independent of $m$, we therefore have proved the following proportionality relations:
\begin{align}
    T_g & \propto m, \label{frag_gl}\\
     T_m & \propto m. \label{frag}
\end{align}

At this point, for the crystal melting we should recall the Lindemann result, Eq. \eqref{constant} \cite{Granato,ubbelohde1965melting}:
\begin{equation}
     T_m \propto 1/\alpha.  \label{alpha_x}
\end{equation}
We should also consider the extension of this relationship to  glasses, in particular the quantitative relation between the glass transition temperature and the thermal expansion coefficient derived based on the molecular-level description of nonaffine displacements in Ref. \cite{ZacconePRL,Anzivino2023}. With all its physical parameters, the relationship reads as \cite{ZacconePRL}:
\begin{equation} 
T_{g} =\frac{1}{\alpha} (1- C - \phi_c^{*} +2\Lambda) - \frac{2\Lambda}{\alpha \, n},\label{AZET}
\end{equation}
where, for polymers, $n$ is the degree of polymerization (i.e. the average number of repeating subunits in a molecule), $C$ ($ \approx 0.48$ for polystyrene glass) is a constant which appears in the relation between molecular packing fraction at the glass transition $\phi_c^* \approx 0.64$ \cite{stachurski,bonn,Zaccone2022,Likos} and temperature $T$, $\ln \big( 1/ \phi \big) = \alpha T + C$. Furthermore, $\Lambda$ is a constant of order $0.1$ that appears in the relation between the molecular packing fraction and the average number of covalent bonds per molecule, $z_\textrm{co}$: $\phi_c = \phi_c^* - \Lambda \cdot z_\textrm{co}$ \cite{ZacconePRL}.

By combining Eqs. \eqref{frag_gl}-\eqref{frag} and Eqs. \eqref{alpha_x}-\eqref{AZET}, we thus obtain:
\begin{align}
    T_g & \propto m/\alpha, \\
     T_m & \propto m/\alpha,  \label{alpha}
\end{align}
which, to our knowledge, is the first theoretical derivation of the universal relations discovered in experimental data for a hundred different materials in Refs. \cite{Lunkenheimer2023} and \cite{Lunk_2024}.

\section{Experimental and computational determination of $G_{NA}$}
The nonaffine correction $G_{NA}$ is harder to measure experimentally, compared to the high-frequency shear modulus $G_A$. However, this would be possible e.g. in rheometry or dynamic mechanical analysis (DMA) by applying a frequency scan to the material over a broad frequency range: e.g. from 0.1 Hz where $G\approx G_{A} - G_{NA}$ to $10^6$ Hz where $G \approx G_{A}$. Alternatively, one can combine different experimental techniques: for soft materials, one can use DMA or standard rheometry at very low frequencies to get $G=G_{A} - G_{NA}$, and a different technique to extract $G_A$ at much higher frequencies, e.g. via nanoindentation \cite{Christoefl} or via Brillouin scattering which probes the shear modulus at frequencies of the order of GHz \cite{Elsayad}. For hard materials, one can apply the same strategy, and extract the low-frequency $G$ from standard mechanical testing, and the high-frequency $G_A$ from, again, AFM-nanoindentation \cite{Wagner2011} or ultrasound measurements \cite{Johnson}. The experimental estimate of $G_{NA}$ can then be obtained as $G_{NA} \approx G_{A} - G$.

We also note that the above derivation provides a new interpretation of the so-called Arrhenius crossover temperature, in liquids, i.e. the temperature that marks a crossover from the high-temperature Arrhenius behaviour of viscosity to the cooperative, typically non-Arrhenius temperature dependence of diffusivity and viscosity in the supercooled liquid \cite{YZ}. It was found, indeed, that $m \sim T_g/T_A$ \cite{YZ}. Comparing this with Eq. \eqref{ket} derived above we get: $T_g = \left(\frac{\ln 10}{c\, G_{NA}(T_g)}\right) m$, which leads to 
\begin{equation}
  T_A \sim 1/G_{NA}(T_g).
\end{equation}
This new relationship derived thanks to the model of our paper is physically significant: it says that the crossover from Arrhenius to cooperative dynamics will occur at higher temperatures when the nonaffine contribution, leading to the vanishing of the shear modulus at $T_g$, is smaller. This is meaningful because $G_{NA}$ is smaller for systems with a higher degree of centrosymmetry \cite{Milkus}, hence with a higher degree of cooperativity, with the latter vanishing only at a higher  temperature $T_A$.

\section{Conclusions}
In summary, we have presented a mathematical derivation of the relationship between melting temperature $T_m$ and glass transition temperature $T_g$, and we also derived expressions for the two temperatures as functions of the fragility of the liquid and of the thermal expansion of the solid. The derivation is crucially based on the generalized Born melting criterion \cite{Born_melting} for both crystals and glasses, suitably extended to include nonaffine displacements at the atomic/molecular level \cite{Zaccone_book}. In particular, without accounting for the nonaffine correction to the shear modulus at melting or at the glass transition, it would be impossible to retrieve the direct proportionality relation between of $T_g$ and $T_m$ with the liquid fragility index $m$. This is the most important result of this paper, and cannot be derived by other theories of melting or of the glass transition, to the best of our knowledge.
Since the fragility $m$ can be related to details of the short-range part of the pair correlation function and to details of the interatomic potential \cite{KSZ}, while the thermal expansion coefficient can be related to the attractive longer-range part of the pair interaction \cite{Kittel}, the above results may open the way for the chemical design of systems where $T_m$ and $T_g$ can be tuned by means of the interparticle interactions and structure \cite{Truskett_inverse}.

\section*{Data availability}
Data sharing is not applicable to this article as no new data were created or analyzed in this study.

\subsection*{Acknowledgments}
K.S. gratefully acknowledges discussions with P. Lunkenheimer and A. Loidl.
A.Z. gratefully acknowledges funding from the European Union through Horizon Europe ERC Grant number: 101043968 ``Multimech'', from US Army Research Office through contract nr. W911NF-22-2-0256, and from the Nieders{\"a}chsische Akademie der Wissenschaften zu G{\"o}ttingen in the frame of the Gauss Professorship program. The authors would like to thank Reviewer 2 for suggesting the connection with the Arrhenius crossover temperature.

\bibliographystyle{apsrev4-1}

\bibliography{refs}

\end{document}